\documentclass[aps,prl,twocolumn,superscriptaddress]{revtex4-1}
\usepackage{graphicx}
\usepackage{color}
\usepackage{amsfonts}
\usepackage{isomath}
\usepackage{amsmath}
\usepackage{amsthm}
\usepackage{amsfonts}
\usepackage{braket}

\usepackage{hyperref}

\usepackage{yhmath}
\usepackage{gensymb}

\def\WS2{WS$_2$}
\def\MoS2{MoS$_2$}

\begin{document}

\title{Spectroscopic view of ultrafast charge carrier dynamics in  single- and bilayer transition metal dichalcogenide semiconductors}

\author{Paulina Majchrzak}
\author{Klara Volckaert}
\affiliation{Department of Physics and Astronomy, Interdisciplinary Nanoscience Center, Aarhus University,
8000 Aarhus C, Denmark}
\author{Antonija Grubi\v{s}i\'{c} \v{C}abo}
\affiliation{Department of Applied Physics, KTH Royal Institute of Technology, Hannes Alfv\'ens v\"ag 12, 114 19 Stockholm, Sweden}
\author{Deepnarayan Biswas}
\author{Marco Bianchi}
\affiliation{Department of Physics and Astronomy, Interdisciplinary Nanoscience Center, Aarhus University,
8000 Aarhus C, Denmark}
\author{Sanjoy K. Mahatha}
\affiliation{Ruprecht Haensel Laboratory, Deutsches Elektronen-Synchrotron DESY, 22607 Hamburg, Germany}
\author{Maciej Dendzik}
\affiliation{Department of Applied Physics, KTH Royal Institute of Technology, Hannes Alfv\'ens v\"ag 12, 114 19 Stockholm, Sweden}
\author{Federico Andreatta}
\author{Signe S. Gr{\o}nborg}
\affiliation{Department of Physics and Astronomy, Interdisciplinary Nanoscience Center, Aarhus University,
8000 Aarhus C, Denmark}
\author{Igor Markovi\'c}
\affiliation{SUPA, School of Physics and Astronomy, University of St Andrews, St Andrews KY16 9SS, United Kingdom}
\affiliation{Max Planck Institute for Chemical Physics of Solids, N\"othnitzer Str. 40, 01187 Dresden, Germany}
\author{Jonathon M. Riley}
\affiliation{SUPA, School of Physics and Astronomy, University of St Andrews, St Andrews KY16 9SS, United Kingdom}
\author{Jens C. Johannsen}
\affiliation{Institute of Condensed Matter Physics, \'Ecole Polytechnique F\'ed\'erale de Lausanne (EPFL), Switzerland}
\author{Daniel Lizzit}
\affiliation{DPIA, University of Udine, Via delle Scienze 206, 33100 Udine, Italy}
\author{Luca Bignardi}
\affiliation{Department of Physics, University of Trieste, Via Valerio 2,Trieste 34127, Italy}
\author{Silvano Lizzit}
\affiliation{Elettra-Sincrotrone Trieste S.C.p.A., S.S. 14 km 163.5, 34149 Trieste, Italy}
\author{Cephise Cacho}
\affiliation{Diamond Light Source, Didcot, OX110DE, U.K.}
\author{Oliver Alexander}
\author{Dan Matselyukh}
\author{Adam S. Wyatt}
\author{Richard T. Chapman}
\author{Emma Springate}
\affiliation{Central Laser Facility, STFC Rutherford Appleton Laboratory, Harwell, United Kingdom}
\author{Jeppe V. Lauritsen}
\affiliation{Department of Physics and Astronomy, Interdisciplinary Nanoscience Center, Aarhus University,
8000 Aarhus C, Denmark}
\author{Phil D. C. King}
\affiliation{SUPA, School of Physics and Astronomy, University of St Andrews, St Andrews KY16 9SS, United Kingdom}
\author{Charlotte E. Sanders}
\author{Jill A. Miwa}
\author{Philip~Hofmann}
\author{S{\o}ren Ulstrup}
\email{ulstrup@phys.au.dk}
\affiliation{Department of Physics and Astronomy, Interdisciplinary Nanoscience Center, Aarhus University,
8000 Aarhus C, Denmark}

\begin{abstract}
The quasiparticle spectra of atomically thin semiconducting transition metal dichalcogenides (TMDCs) and their response to an ultrafast optical excitation critically depend on interactions with the underlying substrate. Here, we present a comparative time- and angle-resolved photoemission spectroscopy (TR-ARPES) study of the transient electronic structure and ultrafast carrier dynamics in the single- and bilayer TMDCs \MoS2 and \WS2 on three different substrates: Au(111), Ag(111) and graphene/SiC. The photoexcited quasiparticle bandgaps are observed to vary over the range of 1.9-2.3~eV between our systems. The transient conduction band signals decay on a sub-100~fs timescale on the metals, signifying an efficient removal of photoinduced carriers into the bulk metallic states. On graphene, we instead observe two timescales on the order of 200~fs and 50~ps, respectively, for the conduction band decay in MoS$_2$. These multiple timescales are explained by Auger recombination involving MoS$_2$ and in-gap defect states. In bilayer TMDCs on metals we observe a complex redistribution of excited holes along the valence band that is substantially affected by interactions with the continuum of bulk metallic states.\\

\end{abstract}

\maketitle

\section{Introduction}
Semiconducting transition metal dichalcogenides (TMDCs) in the 2$H$ structural modification with the formula unit MX$_2$ (M = $\{$Mo, W$\}$; X = $\{$S, Se$\}$) have attracted sustained attention due to their indirect-to-direct bandgap crossover upon thinning to the single layer (SL) limit \cite{Mak:2010,Zhao:2013,Zhang:2014}. 
The bandstructure in this class of materials can additionally be externally tuned via strain \cite{Feng:2012,Johari:2012}, as well as doping via alkali adsorption \cite{miwa2:2015,ZhangX:2016,Kang:2017,Katoch:2018}, electrostatic gating \cite{Ramasubramaniam:2011,Ross:2013,Chernikov:2015,Nguyen:2019} and substrate interactions \cite{Ugeda:2014,Bruix:2016,Raja:2017,UlstrupNano:2019,Waldecker:2019}. Compared to conventional bulk semiconductors, SL TMDCs exhibit exceptionally large exciton and trion binding energies due to reduced dielectric screening of Coulomb interactions in the 2D material \cite{Mak:2010,Mak:2012,Chernikov:2014,Ugeda:2014,Ye:2014,Raja:2017}. 
Moreover, broken inversion symmetry in the trigonal prismatic unit cell of a SL provides a possibility for spin-selective excitation of carriers around the direct bandgap at the $\bar{\mathrm{K}}$ and $\bar{\mathrm{K}}^{\prime}$ valleys of the materials using circularly polarised light \cite{Xiao:2012,Mak:2014,Ulstrup:2017}. The high degree of control over the electronic properties and strong light-matter interactions elevate semiconducting TMDCs to promising candidates for realizing novel applications in photonics, optoelectronics and spintronics \cite{Wang:2012,Xiaodong:2014,Makp:2016,Liang:2020}. 

Successfully integrating TMDCs with such applications requires a complete understanding of interactions with the underlying substrate and precise control of the quasiparticle and free carrier dynamics induced by an optical excitation. It has been shown that metallic substrates strongly influence the bandstructure of the TMDCs. For example, an insulator-to-metal phase transition of SL \WS2 and \MoS2 has been observed on Ag(111) due to hybridization of TMDC bands with the underlying metallic states \cite{Dendzik:2017,DoAmaral:2021,Blue:2020}. A significant substrate-induced bandgap renormalization has been observed in the TMDCs due to dielectric screening \cite{Ugeda:2014,Bruix:2016,Raja:2017}. The optical pulse itself can also lead to bandgap renormalization by creating a large population of strongly screening free carriers \cite{ChernikovNP:2015,Pogna:2016,Ulstrup:2016}. Furthermore, the optoelectronic properties of TMDCs are sensitive to defects, which may be intentionally or inadvertently introduced during or post synthesis \cite{Gutierrez:2013}, and which play a role in the ultrafast dynamics \cite{Wang:2015,Li:2019,Li:2020,Liu:2020}.

Time- and angle-resolved photoemission spectroscopy (TR-ARPES) has provided important insights to the energy- and momentum-dependent ultrafast carrier dynamics of TMDCs, directly revealing renormalization of the quasiparticle bandgap and the associated free carrier dynamics in SL MoS$_2$ \cite{Grubisic:2015,Ulstrup:2016}, as well as interlayer carrier injection in a heterostructure composed of SL WS$_2$ and graphene \cite{Aeschlimann:2020}. Control of spin- and valley-dynamics has been demonstrated via circular dichroism in polarization dependent TR-ARPES on SL WS$_2$ \cite{Ulstrup:2017,Beyer:2019}. In contrast, interlayer interactions in bilayer (BL) MoS$_2$ have lead to the visualization of a momentum-dependent linear dichroism effect \cite{Volckaert:2019}. Furthermore, layer-  and valley-selective optical excitations and intervalley scattering processes have been studied by TR-ARPES on bulk TMDCs \cite{Hein:2014,Wallauer:2016,Bertoni:2016,Wallauer:2020}.

Here, we present a comparative TR-ARPES study of single- and bilayers of \MoS2 and \WS2 on Au(111), Ag(111) and graphene/SiC substrates. We combine a new analysis of our previous TR-ARPES measurements of SL MoS$_2$/Au(111) \cite{Grubisic:2015}, SL MoS$_2$/graphene \cite{Ulstrup:2016}, SL WS$_2$/Ag(111) \cite{Ulstrup:2017} and BL MoS$_2$/Ag(111)  \cite{Volckaert:2019} with new TR-ARPES results on BL WS$_2$/Au(111). We contrast the ultrafast dynamics in SL \MoS2 on the graphene substrate with samples supported on the metallic substrates. The photoexcited quasiparticle bandgaps are compared across these systems, demonstrating an overall variation in gap size of 0.4~eV. The impact on the carrier dynamics of shifting the valence band maximum (VBM) from $\bar{\mathrm{K}}$ in a SL to $\bar{\mathrm{\Gamma}}$ in a BL and the associated interplay with bulk metallic states are determined, resulting in a complex picture of excited hole scattering in BL TMDCs supported on metals. Our work emphasizes the strong substrate dependence of carrier dynamics in SL and BL TMDCs from the perspective of TR-ARPES, providing an outset for expanding this methodology to advanced heterostructures and devices based on the TMDCs.

\section{Experimental}

The synthesis of our TMDC SL and BL samples generally follows the procedure of evaporating Mo or W onto a given substrate in ultra-high vacuum, followed by annealing in a low background pressure of H$_2$S. For details we refer to Ref. \citenum{Miwa:2015} (MoS$_2$/graphene), Refs. \citenum{Gronborg:2015,Bana:2018,Volckaert:2019} (MoS$_2$ on Au(111) and Ag(111)) and Refs. \citenum{Dendzik:2015,Dendzik:2017} (WS$_2$ on Au(111) and Ag(111)).

\begin{figure*}[t!] 
\begin{center}
\includegraphics[width=0.98\textwidth]{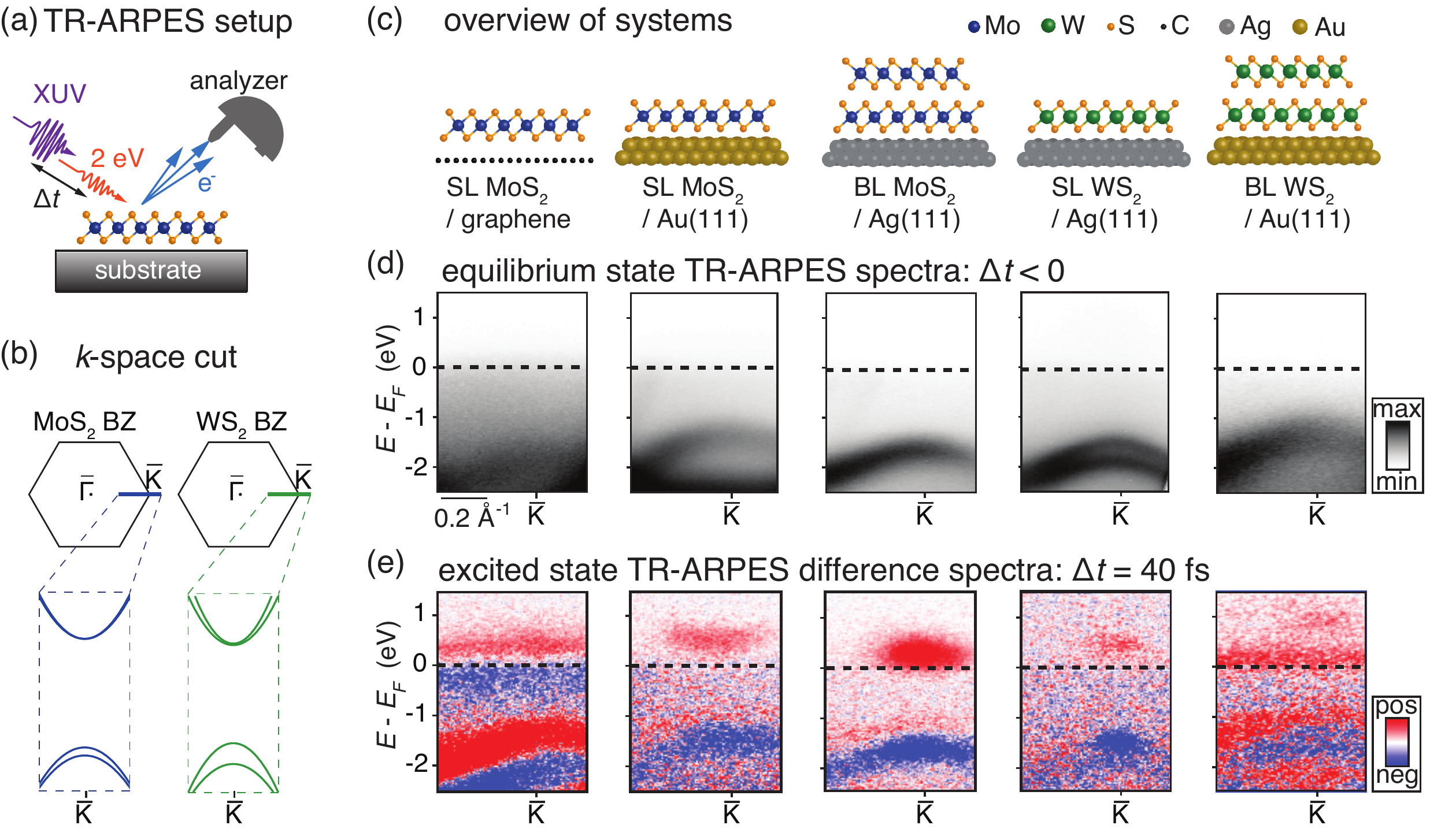}
\caption{(a) Sketch of TR-ARPES setup. (b) Brillouin zone (BZ) with TR-ARPES $k$-space cut indicated by blue and green lines for MoS$_2$ and WS$_2$, respectively. The cut provides access to the VB and CB dispersion around the direct gap of MoS$_2$ and WS$_2$, sketched within the dashed boxes. (c) Side-view diagrams of the systems studied in this work. (d) Photoemission intensity obtained for each system in equilibrium conditions before arrival of the pump pulse ($\Delta t < 0$). (e) Photoemission intensity difference at the peak of optical excitation ($\Delta t = 40$~fs). Each spectrum in (d)-(e) was obtained around $\bar{\mathrm{K}}$ and corresponds to the system sketched in the same column in (c). The position of $E_F$ is indicated by a horizontal dashed line.}
\label{fig:1}
\end{center}
\end{figure*}

The TR-ARPES measurements were performed at Artemis at the Central Laser Facility, Rutherford Appleton Laboratory. The experimental setup is presented schematically in Fig. \ref{fig:1}(a). All the samples were optically excited with a 2 eV pump pulse, which is near-resonant with the quasiparticle bandgaps of our systems. The fluence of the pump beam was kept at approximately 3.0 mJ/cm$^{2}$ in order to maximize the pump-probe
signal while avoiding space-charge effects \cite{ULSTRUP2015340}. All SL samples were probed with 25-eV extreme ultraviolet (XUV) pulses, while the BL samples were probed with 32.5-eV pulses.  The probe pulses in all cases were produced via high harmonic generation in an argon gas jet. Both pump and probe beams were kept linearly polarized in these measurements. The time, energy and angular resolution were approximately 40 fs, 400 meV and 0.2\degree, respectively. The photoemission intensity was collected along the $\bar{\mathrm{\Gamma}}$-$\bar{\mathrm{K}}$ high-symmetry direction of the hexagonal Brillouin zone (BZ), providing access to the dispersion around the direct bandgap at $\bar{\mathrm{K}}$ of MoS$_2$ and WS$_2$, as sketched in Fig. \ref{fig:1}(b). 

\section{Results and discussion}

\subsection{Snapshots of excited free carriers in MoS$_2$ and WS$_2$}
We first survey the energy- and momentum-dependent changes to the photoemission intensity around $\bar{\mathrm{K}}$ following optical excitation, comparing the systems sketched in Fig. \ref{fig:1}(c). TR-ARPES spectra measured in equilibrium conditions before optical excitation ($\Delta t < 0$) are shown in Fig. \ref{fig:1}(d). The changes following optical excitation are presented in Fig. \ref{fig:1}(e) via the difference in photoemission intensity between the equilibrium spectra in Fig. \ref{fig:1}(d) and the corresponding spectra taken shortly after the arrival of the pump pulse ($\Delta t = 40$~fs). The red/blue contrast corresponds to gain/loss signal, which stems from three primary effects: (i) Filling/depletion of excited electrons/holes in the associated states, (ii) rigid band shifts caused by renormalization of the bandgap, and (iii) linewidth broadening of the bands that reflects the optically induced change of intrinsic quasiparticle lifetime \cite{Ulstrup:2015,Na:2019,Majchrzak:2020}.

The intensity difference for SL \MoS2/graphene is dominated by the loss signal (blue) in the valence band (VB) accompanied by a stark gain signal (red) directly above it, indicating a rigid shift of the band position in addition to the expected depletion.  A flat region of intensity depletion and gain appears pinned to $E_F$. This suggests that the excitation leads to a significant population of in-gap states (IGS) that arise from defects in the sample. The intensity difference in the IGS above $E_F$ is mixed with that of the finite population of free carriers in the conduction band minimum (CBM) \cite{Ulstrup:2016}. 

For SL MoS$_2$/Au(111), BL MoS$_2$/Ag(111) and SL WS$_2$/Ag(111), the difference spectra primarily display a localized gain of intensity in the conduction band (CB) and a corresponding loss in the VB around $\bar{\mathrm{K}}$, consistent with the presence of excited free electrons and holes in these states. A faint gain signal immediately adjacent to the central loss region in the VB could emerge from either a minute rigid shift of the bands or a slight broadening effect caused by the optical excitation. 

The intensity difference of BL WS$_2$/Au(111) appears more complex than for the other systems. In particular, the system displays a flat gain signal directly above $E_F$, indicating a prominent filling of IGS. Further above $E_F$, a localized gain signal is indicative of the local CBM of BL WS$_2$ at $\bar{\mathrm{K}}$. The substantial gain/loss signal surrounding the VB indicates a more pronounced linewidth broadening in this system.

\begin{figure}[t!] 
\begin{center}
\includegraphics[width=0.49\textwidth]{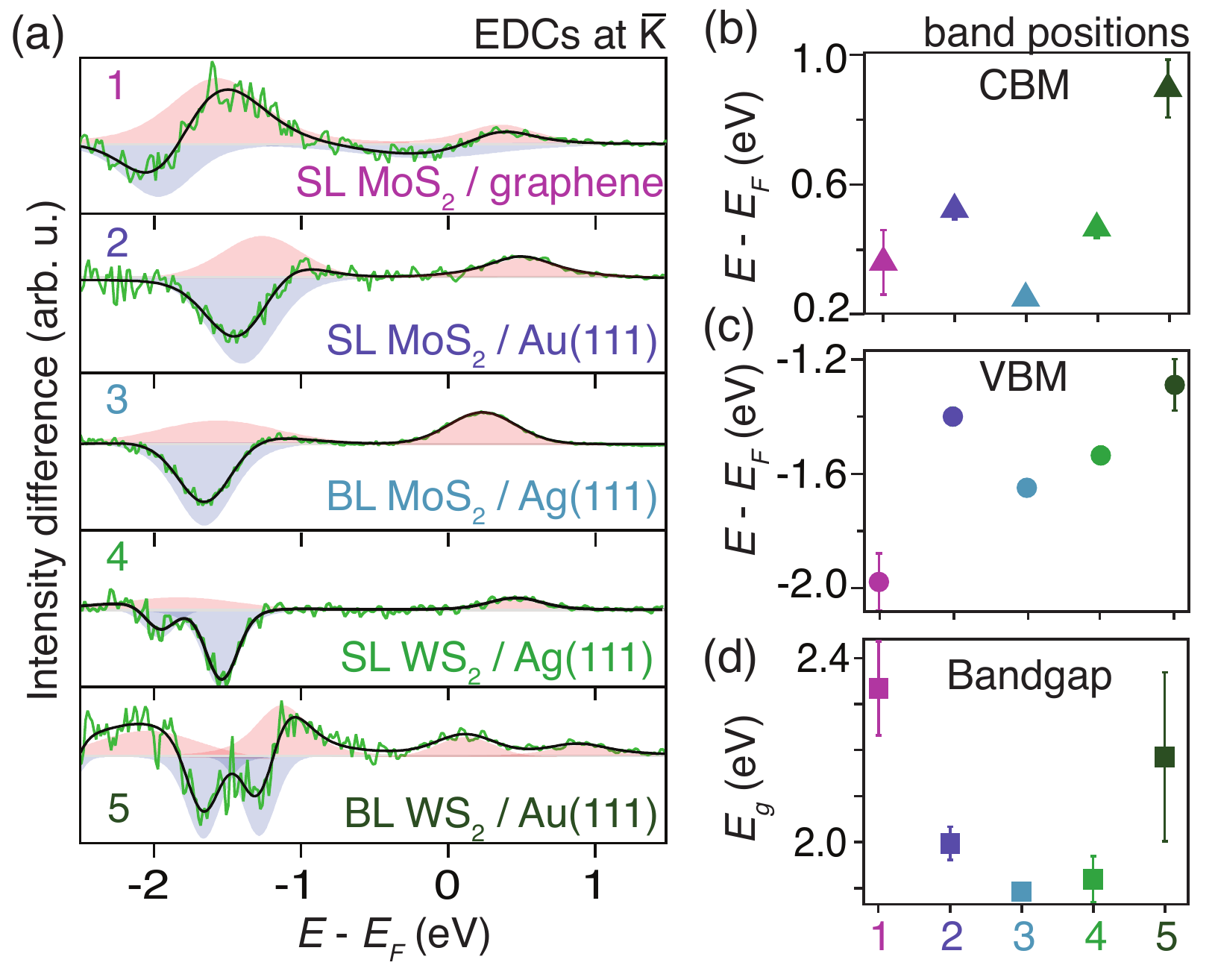}
\caption{(a) EDCs of the intensity difference (green curves) extracted by integrating the photoemission intensity difference over a region of $\pm$0.1 \AA$^{-1}$ around $\bar{\mathrm{K}}$. The smooth curves represent fits to a multi-peak Voigt function. Fitted peak components are shown via filled red (intensity gain) and blue (depletion) areas. (b)-(c) Positions of (b) CBM and (c) VBM extracted from the EDC analysis in (a). (d) Resulting direct bandgap at $\bar{\mathrm{K}}$. The numbering of systems on the horizontal axis of (b)-(d) is referenced to the five systems in (a).}
\label{fig:2}
\end{center}
\end{figure}

\subsection{Extraction of photoexcited quasiparticle bandgaps}

We obtain a quantitative estimate of the photoexcited direct bandgap, $E_g$, at $\bar{\mathrm{K}}$ by analyzing energy distribution curves (EDCs) of the intensity difference, binned $\pm 0.1$~\AA$^{-1}$ around $\bar{\mathrm{K}}$. The EDCs are presented via green curves in Fig. \ref{fig:2}(a). A multi-peak fit to Voigt functions (smooth curves) permit a decomposition of the EDCs into single component peaks, as shown via filled red (blue) areas that signify gain (loss) in the photoemission signal. These are referred to as gain and loss peaks in the following discussion. We choose to analyze EDCs of the intensity difference instead of EDCs obtained from the raw spectra, as the complex background signal, that strongly varies between the systems (see Fig. \ref{fig:1}(d)), is removed in the difference spectra. This introduces a small error of maximum 40 meV when comparing the extracted band position of the difference and raw EDCS

In the VB region, the EDC for SL \MoS2/graphene displays negative and positive peaks adjacent to each other. This is caused by a rigid energy shift due to a bandgap renormalization induced by the photoinduced free carriers \cite{Steinhoff:2014,Ulstrup:2016}. The VBM extracted from the equilibrium spectra coincides with the loss peak minimum. Around $E_F$, the larger amplitude of the gain peak compared to the loss peak is interpreted as a sign of the IGS superimposed on the excited CBM signal. We assume that the CBM coincides with the position of the gain peak maximum above $E_F$. For the SL TMDCs and BL MoS$_2$ on metallic substrates, the isolated gain peaks above $E_F$ clearly indicate the CBM energies. The VBM energies are obtained from the corresponding loss peaks below $E_F$. The slight broadening that is observed around the VB for these systems  in Fig. \ref{fig:1}(e) is accounted for in the fit, as evidenced by the broad gain peaks in the vicinity of the VB loss peaks. In BL WS$_2$/Au(111), the substantial linewidth broadening of the VB is best described by two gain peaks adjacent to the central VB loss peaks. Furthermore, the two gain peaks above $E_F$ in this system permit us to separate the IGS and CBM by their positions in energy, IGS being the closest to $E_F$, for the further analysis. The two adjacent VB loss peaks observed for SL and BL WS$_2$ reflect the spin-split VB states that are separated by 420~meV \cite{Dendzik:2015}.

The fitted peak positions corresponding to the CBM and VBM are presented in Figs. \ref{fig:2}(b)-(c). For all the investigated samples, $E_F$ is situated closer to the CBM than the VBM, indicating $n$-type doping. This is consistent with theoretical work, where the doping is attributed to the metal work function modification upon charge redistribution at the interface as well as the appearance of gap states of predominantly metallic $d$-orbital character \cite{Popov:2012,Gong:2014}. 
We note that hybridization with substrate states can also shift and distort bands \cite{Bruix:2016,Shao:2019}. The VB at $\bar{\mathrm{K}}$ is situated at a higher binding energy for \MoS2 and \WS2 samples supported on Ag(111) compared to their counterparts supported on Au(111), which is a result of Fermi level pinning due to different substrate work functions \cite{Dendzik:2017}. In SL MoS$_2$/graphene and BL WS$_2$/Au(111) the Fermi level is pinned by the IGS.

In Fig. \ref{fig:2}(d), we show the size of the quasiparticle bandgaps determined from the VBM and CBM energies. For the \MoS2 samples, the value of the bandgap is slightly overestimated since we are unable to resolve the spin-splitting of 145 meV \cite{miwa2:2015}, moving the negative intensity difference peak to the middle of the split bands.
The quasiparticle bandgap of $\approx$2.3~eV for the quasi-freestanding SL \MoS2 on graphene is significantly larger than that for samples on metallic supports due to weaker dielectric screening and absence of hybridization between TMDC and substrate electronic bands. In optically excited SL \MoS2/graphene, the bandgap is reduced by the free carrier density, which shifts the CB and VB towards each other \cite{Ulstrup:2016}. For TMDCs on metal supports, the bandgaps are substantially renormalized by the strong dielectric screening from the substrate, reaching a minimum of $\approx$1.9~eV. Interestingly, we observe a somewhat larger direct bandgap of $\approx$2.2~eV in BL \WS2/Au(111). This may be explained by the global minimum of the CB shifting away from $\bar{\mathrm{K}}$ to the $\bar{\mathrm{\Sigma}}$ valley between $\bar{\mathrm{\Gamma}}$ and $\bar{\mathrm{K}}$ in BL WS$_2$ \cite{Zeng:2013}, pushing the energy of the CB states at $\bar{\mathrm{K}}$ further from the VB. 

\subsection{Analysis of carrier relaxation timescales}

\begin{figure*}[t!]  
\begin{center}
\includegraphics[width=0.7\textwidth]{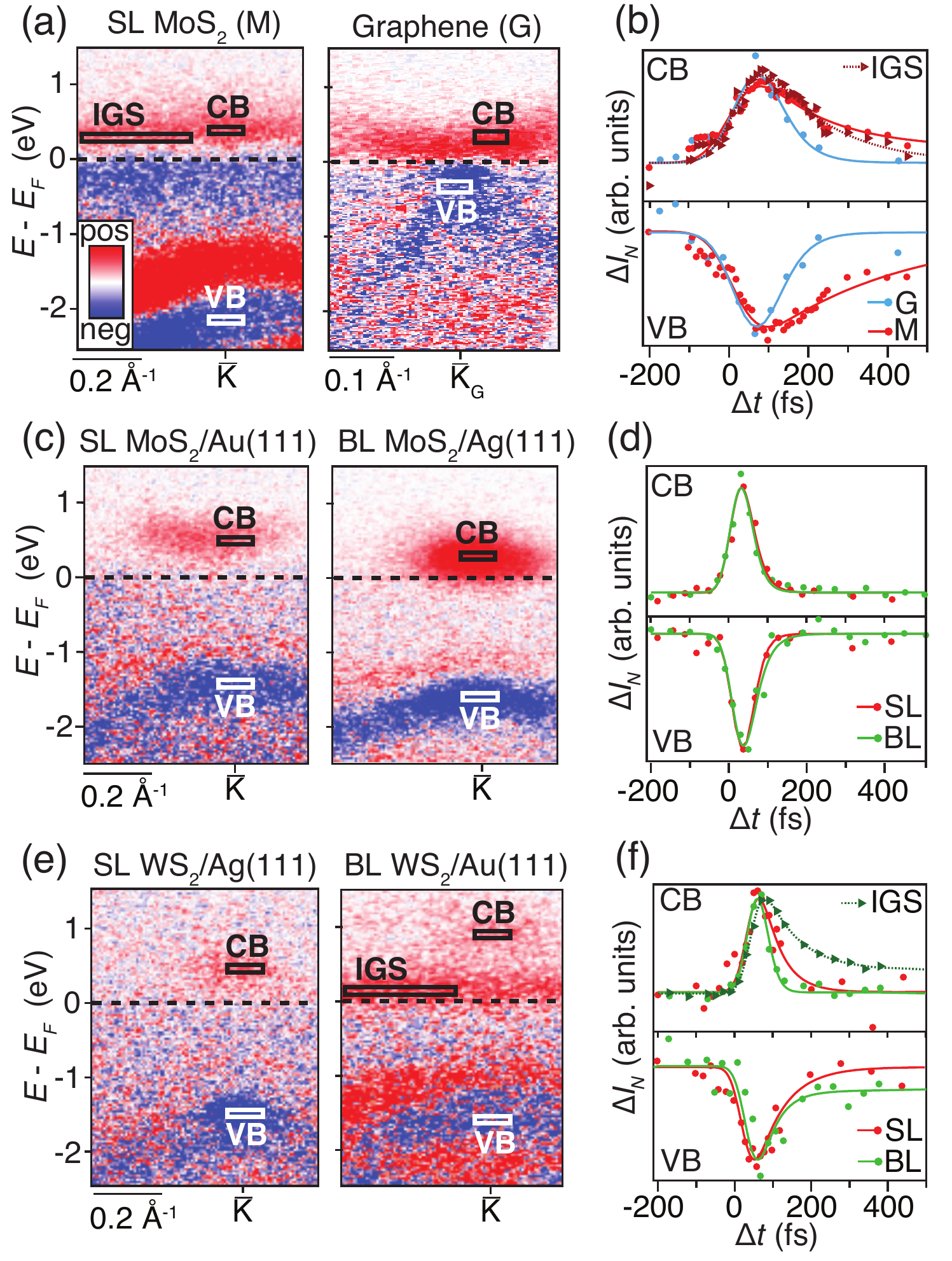}
\caption{(a) Intensity difference at the peak of excitation for (left) SL MoS$_2$/graphene and (right) underlying graphene around the Dirac point $\bar{\mathrm{K}}_G$. (b) Integrated photoemission intensity difference (markers) as a function of time within the corresponding boxed regions shown in (a).  Smooth curves represent fits to exponential functions convoluted with a Gaussian function that accounts for the temporal resolution. The dynamics in the CB and IGS of SL \MoS2/graphene are described by biexponential functions, while the other regions are described by single exponentials. (c)-(f) Similar analysis as in (a)-(b) presented for (c)-(d)  SL MoS$_2$/Au(111) and BL MoS$_2$/Ag(111) and (e)-(f) SL WS$_2$/Ag(111) and BL WS$_2$/Au(111).  In (f), the dynamics in the VB and IGS of BL \WS2/Au(111) are described by biexponential functions while the remaining dynamics are described by single exponentials.}
\label{fig:3}
\end{center}
\end{figure*}

The time-dependent evolution of the excitation and ensuing recombination processes in our systems is investigated by integrating the normalized intensity difference within boxes around the $(E,k)$-regions containing the CBM and VBM at $\bar{\mathrm{K}}$, as well as an IGS region away from $\bar{\mathrm{K}}$, as shown for each system in Figs. \ref{fig:3}(a), \ref{fig:3}(c) and \ref{fig:3}(e). The resulting time-dependence of integrated intensity difference, denoted by $\Delta I_N$, is shown for each system in Figs. \ref{fig:3}(b), \ref{fig:3}(d) and \ref{fig:3}(f). The characteristic decay time constants of the signals for each system are extracted by fitting with exponential functions convoluted with a Gaussian temporal resolution function. The resulting decay constants for the CBM, VBM and IGS regions are summarized in Tab. \ref{tab:1} and discussed further below.

\begin{table*} [t!]  
  \centering
  \begin{tabular*}{0.8\textwidth}{@{\extracolsep{\fill} } l c c c c c}
  \hline		
    \hline	
   System & $E_g$ (eV)  & $\tau_{\mathrm{CB}}$ (fs) & $\tau_{\mathrm{VB}}$ (fs)& $\tau_{\mathrm{IGS}}$ (fs) & $\tau_2$ (fs)\\
  \hline	
  SL MoS$_2$/graphene		&	2.33 $\pm$ 0.10 &	171 $\pm$ 28		&		355 $\pm$ 37		&		157 $\pm$ 23 	&		$\sim$50$\cdot 10^3$ \\
  SL MoS$_2$/Au(111)	& 2.00 $\pm$	 0.04 &	33 $\pm$ 20		&		30 $\pm$ 20 	&		-- 	&		-- \\
  BL MoS$_2$/Ag(111)	&	1.89 $\pm$ 0.01 &	34 $\pm$ 20		&		30 $\pm$ 20 	&		-- &		--  \\
  SL WS$_2$/Ag(111)	&	1.92 $\pm$ 0.05 &	49 $\pm$  20		&		72 $\pm$ 34 	&		--  &		--  \\
  BL WS$_2$/Au(111)	& 2.19 $\pm$ 0.18	&	30 $\pm$ 20		&		43 $\pm$ 20 	&		84 $\pm$ 27 	&		$\sim$500  \\
  \hline  
  \hline
\end{tabular*}
  \caption{Relaxation time scales resulting from the fits in Figs. \ref{fig:3}(b), \ref{fig:3}(d) and \ref{fig:3}(f). The quasiparticle bandgaps resulting from the analysis in Fig. \ref{fig:2}(d) have been added for completeness. In SL MoS$_2$/graphene (BL WS$_2$/Au(111)), $\tau_2$ corresponds to the second component of the biexponential decay of CB (VB) and IGS signals. }
   \label{tab:1}
\end{table*}

In SL \MoS2/graphene the VB signal is described by a single exponential function with time constant (355$\pm$37)~fs, while the CB signal is described by a biexponential function with time constants of (171$\pm$28) fs and approx. 50 ps. The very long component is outside our detection window and indicates a strong coupling between the CB and IGS via a slow-acting recombination mechanism that does not involve the VB states  \cite{Wang:2015}. This is supported by the very similar time-dependence of the IGS and CB signals. The time-dependence of the corresponding excitation and relaxation processes in the underlying graphene in the boxes around the Dirac cone at $\bar{\mathrm{K}}_G$, shown in Fig. \ref{fig:3}(a), is inspected in order to determine whether there is any charge transfer between MoS$_2$ and graphene. We observe that the electron and hole signals are highly symmetric in graphene, the rise-time of the signals is identical with what we see in MoS$_2$, and the subsequent relaxation is much faster in graphene. These three observations indicate that we can exclude significant charge transfer between the two systems in our sample. We note that the graphene and MoS$_2$ lattices are predominantly rotated by 30$^{\circ}$ with respect to each other as the result of our growth method \cite{Miwa:2015}, preventing hybridization between the Dirac cone and the MoS$_2$ VBM and CBM states and thereby removing any efficient charge transfer channels.

In MoS$_2$ on metallic substrates and SL WS$_2$ on Ag(111), the recombination dynamics is significantly  faster. For SL and BL \MoS2, the signal is symmetric between VB and CB. It can be described with a single decay constant around 30 fs, which is comparable to our time-resolution. In SL \WS2, the extracted decay time is slightly higher, $\approx$50 fs for the CB and $\approx$70 fs for the VB. These ultrafast relaxation processes proceed via electron-hole pair recombination involving the bulk continuum of states in the metal substrates \cite{Grubisic:2015}. For BL \WS2/Au(111) the excited CB signal displays a similar ultrafast decay. However, the VB and IGS signals are described by biexponential functions with a fast component that is  $\approx$40~fs for the VB and  $\approx$80~fs for the IGS and identical slow components of 500 fs. This suggests a slow recombination between VB states and the IGS in this system, irrespective of the efficient scattering channels available in the metal. One may speculate that this behavior arises from the IGS coming from defects in the top layer of the BL WS$_2$ that are therefore essentially decoupled from the underlying Au(111). 

In all the systems considered here, the decay of photoexcited carriers proceeds several orders of magnitude faster than in large bandgap bulk semiconductors, where the direct band-to-band recombination of the photoexcited carriers occurs on timescales varying from hundreds of picoseconds in GaAs \cite{Othonos:1998}, to tens of nanoseconds in rutile TiO$_2$ \cite{Yamada:2012}.

\subsection{Dynamics along $\bar{\mathbf{\Gamma}}$-$\bar{\mathbf{K}}$ in bilayers}

\begin{figure*}[t!] 
\begin{center}
\includegraphics[width=1\textwidth]{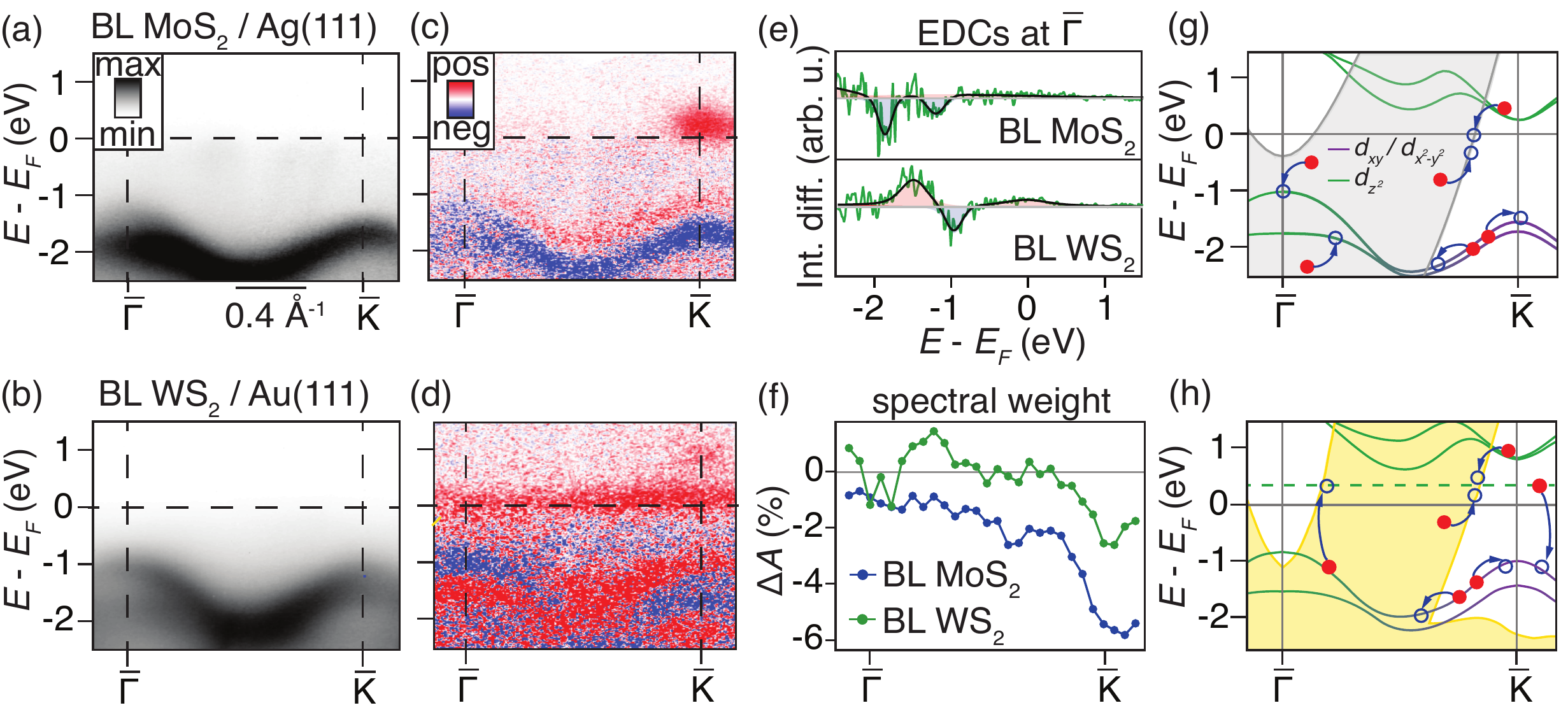}
\caption{ (a)-(b) ARPES intensity of (a) BL MoS$_2$/Ag(111) and (b) BL WS$_2$/Au(111) along $\bar{\mathrm{\Gamma}}$-$\bar{\mathrm{K}}$ before optical excitation. (c)-(d) Corresponding intensity difference between spectra acquired at the peak of excitation ($\Delta t \approx 40$~fs) and the equilibrium spectra in (a)-(b). (e) EDCs (green curves) of the intensity difference obtained at  $\bar{\mathrm{\Gamma}}$ in (c)-(d). Smooth curves represent multi-peak Voigt function fits with fitted peak components shown via filled red (intensity gain) and blue (depletion) areas. (f) Change of spectral weight along the VB between equilibrium and peak excitation. (g)-(h) Schematic representation of possible electron (filled red circles) and hole (empty blue circles) recombination processes. In (g) the BL MoS$_2$ bandstructure is sketched and color-coded according to the orbital character of the bands, based on calculations from Refs. \citenum{Walter:2012,Kormnyos2015}. In (h), the similar sketch of the BL WS$_2$ bandstructure is based on the calculations in Refs. \citenum{Zeng:2013,Kormnyos2015}. The dashed horizontal line indicates the IGS. Gray and yellow shading in (g)-(h) represent the projected bulk bandstructure of Ag(111) and Au(111), respectively, adapted from Refs. \cite{Takeuchi:1991,Dendzik:2017}.}
\label{fig:4}
\end{center}
\end{figure*}

In bilayer TMDCs, the interaction of the out-of-plane orbitals between the two layers causes a bonding-antibonding splitting of the VB states at  $\bar{\mathrm{\Gamma}}$, which ultimately shifts the VBM to $\bar{\mathrm{\Gamma}}$ \cite{Walter:2012,Zeng:2013,Gronborg:2015}. By analyzing TR-ARPES spectra obtained along the full $\bar{\mathrm{\Gamma}}$-$\bar{\mathrm{K}}$ line we are able to evaluate how the shift of the VBM away from $\bar{\mathrm{K}}$ influences carrier dynamics in the BL systems. 

In Figs. \ref{fig:4}(a)-(b), we present ARPES spectra along $\bar{\mathrm{\Gamma}}$-$\bar{\mathrm{K}}$ of BL \MoS2/Ag(111) and BL \WS2/Au(111) taken before the arrival of the pump pulse. The primary distinction between the two systems is that the VBM at $\bar{\mathrm{\Gamma}}$ is clearly visible in \WS2, but faded in \MoS2 as a consequence of photoemission matrix element effects. The corresponding difference spectra between equilibrium and peak excitation spectra are shown in Figs. \ref{fig:4}(c)-(d). EDCs of the intensity difference at $\bar{\mathrm{\Gamma}}$ are presented in Fig. \ref{fig:4}(e), following a similar analysis as for the corresponding EDCs at $\bar{\mathrm{K}}$ in Fig. \ref{fig:2}(a). The VB splitting is evident in BL MoS$_2$ in the intensity difference, leading to a value of $(690 \pm 30)$~meV via our fits of the EDCs. For BL WS$_2$ we determine a value for the splitting of $(680 \pm 20)$~meV from the raw spectrum in Fig. \ref{fig:4}(b). 

We determine the change of spectral weight, $\Delta A$, along the VB by extracting EDCs of the raw TR-ARPES spectra and calculate the change of the area under the VB peak between equilibrium and excited state EDCs. We perform this EDC analysis at each $k$ between $\bar{\mathrm{\Gamma}}$ and $\bar{\mathrm{K}}$ in bins of $\pm$0.1~\AA$^{-1}$. The value $\Delta A$ is proportional to the number of excited holes in a given part of the band \cite{Ulstrup:2016}. The extracted dependence of $\Delta A$ on $k$ is presented in Fig. \ref{fig:4}(f). In BL \MoS2/Ag(111), the signal from excited holes peaks around $\bar{\mathrm{K}}$, and is significantly reduced, but not removed, along the band towards $\bar{\mathrm{\Gamma}}$. This appears rather surprising as one would anticipate the holes to accumulate around the minimum energy state at the VBM at $\bar{\mathrm{\Gamma}}$. A possible explanation involves the change of orbital character of the bands from $\bar{\mathrm{K}}$ to  $\bar{\mathrm{\Gamma}}$ and the location of bulk states of the Ag(111) substrate, which are summarized in Fig. \ref{fig:4}(g). At $\bar{\mathrm{K}}$ the in-plane $d_{xy}/d_{x^2-y^2}$ orbital character of the VB results in weaker coupling to the substrate compared to the out-of-plane $d_{z^2}$ orbital character around $\bar{\mathrm{\Gamma}}$ \cite{Bussolotti:2019}. Moreover, the latter states are fully overlapped in energy and momentum with metallic bulk states, facilitating efficient hybridization between these BL MoS$_2$ states and the substrate. Electron-hole recombination processes involving these states therefore proceed efficiently via Auger processes, as sketched in Fig. \ref{fig:4}(g).

In BL WS$_2$/Au(111), the photoexcited holes in the VB have competing scattering channels available: They can either decay into the metallic states, or they can recombine with the IGS, as seen via the time dependent dynamics in Fig. \ref{fig:3}(f). Furthermore, the energy- and momentum-dependence of the bulk continuum of metallic states in Au(111) leads to a peculiar situation where the states at the VBM of BL WS$_2$ are located in the projected bulk gap of the Au(111) at $\bar{\mathrm{\Gamma}}$, as shown in Fig. \ref{fig:4}(h). This leads to a less efficient coupling with the substrate and a peaked concentration of holes at the top of the VB, which is seen via the loss signals around -1~eV in Figs \ref{fig:4}(d)-(e) and the spikes in the green curve at $\bar{\mathrm{\Gamma}}$ in Fig. \ref{fig:4}(f). The remaining gain signals likely result from the additional IGS scattering channels. 

Our analysis thus reveals a complex dependence of carrier dynamics in BL TMDCs on the interaction with underlying substrate states and the orbital characters of the TMDC VB states. When considering the consequences of this dynamics for the use of the SL semiconductors in devices, the efficient removal of carriers demonstrated on the metal substrates would be advantageous around electrodes in a device based on TDMCs. On the other hand, clean access to the ultrafast dynamics of TMDCs along the entire top-most VB requires substrates without electronic states in this region, such as oxide insulators \cite{Katoch:2016} or hexagonal boron nitride \cite{Katoch:2018}.

\section{Conclusion and Outlook}

In summary, we have determined the influence of graphene, Au(111) and Ag(111) substrates and the presence of IGS on the ultrafast dynamics of SL and BL MoS$_2$ and WS$_2$. On Au(111) and Ag(111), the quasiparticle bandgap is significantly reduced, reaching $\approx$1.9~eV, compared to the value of $\approx$2.3~eV determined on the graphene substrate. In the TMDCs supported on metals we found an ultrafast relaxation of the signal on the order of our 40~fs time-resolution. This behavior reflects the efficient electron-hole pair recombination channels provided by the substrates, although for BL WS$_2$ on Au(111) we found a substantially slower 500~fs VB dynamics caused by IGS. The weakly-interacting graphene was not observed to affect the dynamics in the adjacent SL MoS$_2$, which was more strongly dominated by interactions with IGS that led to biexponential relaxation timescales on the order of 150~fs and 50~ps in the CB states.

In the future, we hope to extend this work to TMDC samples on insulating substrates such as metal oxides \cite{Katoch:2016,Chen:2018}, as well as TMDC-based heterostructures \cite{Gong:2014,Chiu:2015}.
Indeed, TR-ARPES studies on SL semiconducting TMDCs have until recently been limited by flake sizes that are small (on the order of 10 $\mu$m) compared to the typical XUV beam size. As the TMDC growth technology is maturing, synthesis of large, single-orientation \MoS2 and \WS2 flakes has become possible on a range of substrates \cite{Bana:2018,Kastl:2018,Chen:2018}, and a variety of sample infrastructures can be engineered by van der Waals pick-up transfer methods \cite{Guo:2021}.
Furthermore, the application of photoemission electron microscopes to pump-probe experiments \cite{Maklar:2020,Madeo:2020} allows for the combination of time- and momentum-resolved spectra with imagining at the microscale.
A full understanding of the nuanced interfacial interactions on the electronic structure of the TMDCs and their heterostructures will have an invaluable impact for tailoring optoelectronic devices.

\section{Acknowledgement}
We thank Phil Rice and Alistair Cox for technical support during the Artemis beamtimes. We gratefully acknowledge funding from VILLUM FONDEN through the Young Investigator Program (Grant. No. 15375) and the Centre of Excellence for Dirac Materials (Grant. No. 11744), the Danish Council for Independent Research, Natural Sciences under the Sapere Aude program (Grant Nos.  DFF-9064-00057B and DFF-6108-00409). Access to the Artemis Facility was funded by STFC. I.M. acknowledges financial support by the International Max Planck Research School for Chemistry and Physics of Quantum Materials (IMPRS-CPQM). The authors also acknowledge The Royal Society and The Leverhulme Trust.

\section{Conflict of interest}
The authors declare no conflict of interest.

\end{document}